\documentclass[twocolumn,showpacs,amsmath,amssymb,a4]{revtex4}

\usepackage[dvips]{graphicx}
\graphicspath{{../figures_rev/}}

\usepackage{dcolumn}
\usepackage{bm}

\begin{document}

\title{Interplay of plasma-induced and fast thermal nonlinearities in a GaAs-based photonic crystal nanocavity}

\author{Alfredo de Rossi}
\email{alfredo.derossi@thalesgroup.com}
\author{Michele Lauritano}
 \altaffiliation[Also at ]{Engineering Department, University of Ferrara, Italy.}
\author{Sylvain Combri\'{e}}
\author{Quynh Vy Tran}%
\author{Chad Husko}%
 \altaffiliation[Also at ]{Columbia University, New York, U.S.A.}
\affiliation{%
Thales Research and Technology, route d\'epartementale 128, 91767 Palaiseau, France}%

\date{\today}

\begin{abstract}
We investigate the nonlinear response of GaAs-based photonic crystal cavities at time scales which are much faster than the typical thermal relaxation rate in photonic devices. We demonstrate a strong interplay between thermal and carrier induced nonlinear effects. We have introduced a dynamical model entailing two thermal relaxation constants which is in very good agreement with experiments. These results will be very important for Photonic Crystal-based nonlinear devices intended to deal with practical high repetition rate optical signals. 
\end{abstract}

\maketitle

\section{Introduction}
With both a small modal volume and large quality factor, optical microcavities exhibit a greatly enhanced light-matter interaction and a strong non-linear optical response\cite{vahala_nat03,Almeida2004,Yanik04,XuNature2007}. An emerging class of optical microcavities is based on air-clad two-dimensional (2D) Photonic crystals (PCs). A Q-factor greater than $10^6$ was achieved with this technology \cite{kuramochi06,song_05,asano_06} which also allows small modal volumes ($\approx (\lambda/n)^3$ or $10^{-19}m^3$. Moreover, 2D PC technology is a planar technology which is particularly suited for the fabrication of photonic circuits. Therefore, the move towards all-optical processing, long considered impractical, has changed dramatically and new possibilities have been opened \cite{soljacic04}. Impressive experimental demonstrations of low-energy ($fJ$) optical bistability and all-optical switching \cite{notomi05,tanabe_ox05,Yang_07}, wavelength conversion \cite{tanabe06}, optomechanical effects \cite{notomi06} and dynamical control of the cavity 	lifetime \cite{tanabe07} are recent noteworthy achievements.
Most of these results come from a specific technology, i.e. silicon-based air-clad PCs where the nonlinear process involved is two-photon absorption (TPA) followed by plasma-induced and thermally-induced refractive index change. The optical power required is quite small ($\approx 100fJ$) and can be very fast (70 ps by ion implant, as the carrier recombination time result much shorter in these nanostructured devices than in bulk silicon \cite{tanabe05,tanabe07apl}. It has been predicted that, owing to strong light-matter interaction, nonlinear properties can be engineered by introducing nanoparticles\cite{Singh_2008}.

\indent The desire for an even faster response time motivates the research on alternative materials with a strong optical Kerr effect. Much progress has been made in processing chalcogenide crystals with high-Q PC microcavities have been demonstrated recently \cite{Ruan2007}. III-V semiconductors are also good candidates for optical switching as they have two very attractive features, compared with silicon. First, the optimization of the Kerr effect with respect to TPA \cite{AitchisonJQE1997,bristow_GaAs_APL07}. Self phase modulation due to Kerr effect has been demonstrated in 2D PCs recently\cite{Oda2007,oda_APL2008}. An additional feature of III-V semiconductors is the possibility of exploiting a strong nonlinear effect related to absorption saturation in active structures, such as Quantum Wells (QWs) and Quantum Dots (QDs). Fast nonlinear dynamics \cite{Raineri_APL04}, leading to bistability \cite{Yacomotti_APL06} and excitability \cite{Yacomotti_excit_PRL06}, has been observed in InP-based 2D PCs with InAsP QWs which are designed to operate with band-edge modes coupled to off-plane free space beams.  
Low power and very fast (2 ps) switching (15 ps for a complete on/off cycle) has been demonstrated with a Symmetric Mach Zehnder-type all-optical switch made with InAs/AlGaAs quantum dots \cite{nakamura06}.

\indent In this paper we focus on the dynamics of the processes initiated by TPA in GaAs PCs.
Although, ideally, all optical switching requires the Kerr effect, in practice TPA, followed by a carrier induced index change, is still a very attractive approach, because of the relative simplicity of the technology.  In particular, there is no active material, e.g. QDs, or phase matching condition required. Moreover, compared to silicon, the TPA coefficient in GaAs is tenfold higher \cite{Dinu2003} and the non-radiative carrier lifetime in patterned structures can be very short (8 ps) \cite{bristow_GaAs_APL03}. A fast and strong nonlinear response is therefore expected in GaAs PC nanocavities. Very recently, we have demonstrated ultra-fast (6 ps recovery time) and low power ($\approx 150$ fJ) modulation in GaAs Photonic crystal cavities\cite{Husko_APL2008}. We have investigated optical bistability in high-Q ($Q=0.7\times 10^6$) PC cavities on GaAs and reported an ultra-low threshold power ($\mu W$ range) \cite{weidner06,weidner07,combrie2008}. After the submission of this manuscript, memory operation have been demonstrated in a InGaAsP PC cavity\cite{Shinya_08}%
 
In these experiments, only the slow ($\mu s$) regime has been explored, which is dominated by the thermally induced index change. This is also due to the high thermal resistance of the membrane structures compared to, for instance, micro-disks \cite{michael_07} or membranes bonded to a $SiO_2$ cladding \cite{Yacomotti_APL06} and the lower thermal conductivity of GaAs with respect to silicon.

In this paper we investigate the response of PC microcavities at a much faster modulation rate (up to 50 MHz), which is well beyond the typical thermal relaxation time of photonic devices. Under these conditions, the analysis of nonlinear responses, such as bistability, cannot be explained with static models. First of all,  moving to faster time-scales modifies the relative influence of thermal and carrier plasma effects. While at very fast time scales (ps) the dynamics tend to be controlled by the carrier lifetime, we will show that in the range between 1 - 100 ns the dynamics results from the interplay of fast thermal effects and carrier plasma index shift. We introduce a model that incorporates two thermal relaxation constants. This is necessary to explain this dynamics and the fact that, despite the high thermal resistance of PCs microcavities, there are thermal effects which develop in less than 10 ns. Section II is devoted to experiments made on a PC microcavity coupled to a waveguide. In section III we will introduce our multi-scale model.  Discussion of the results forms the body of section IV. 

\section{Experimental setup and sample description}
\begin{figure}[h]
\includegraphics[width=7.5cm]{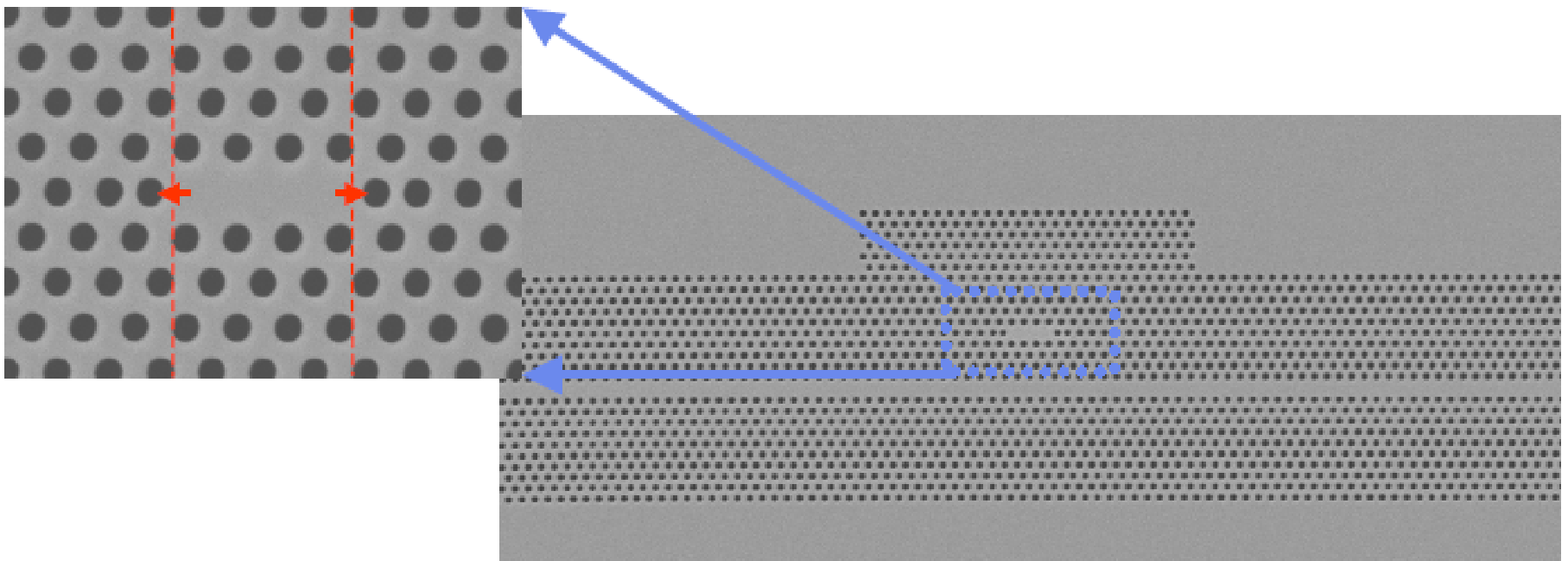}
\includegraphics[width=7.5cm]{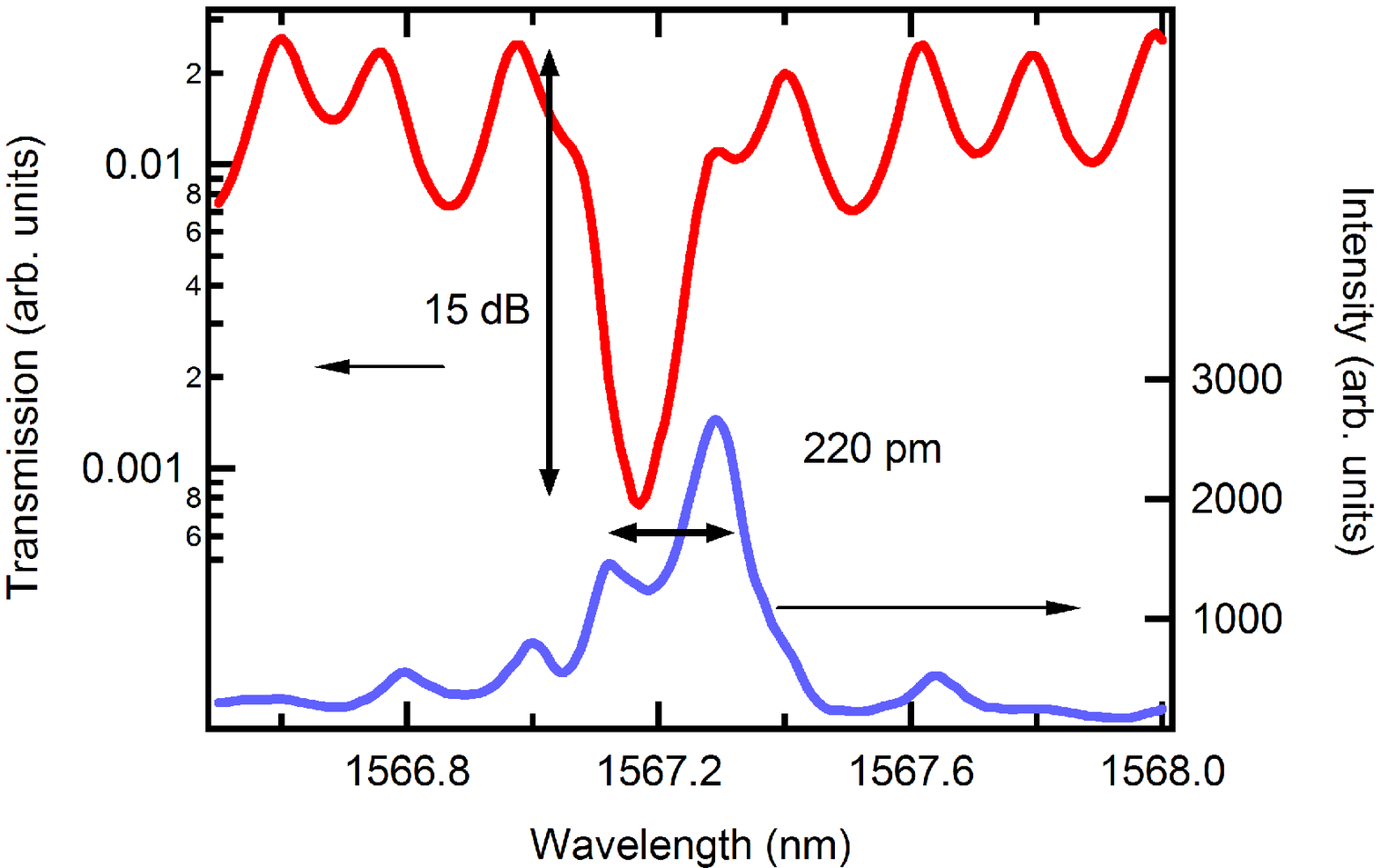}
\caption{(Color online) 	Top: SEM image of the sample, also showing a line-defect photonic crystal waveguide with side coupled $L3$ cavity. Bottom: linear transmission near resonance showing that the cavity is over-coupled. The linewidth of the cavity is larger than the transmission dip because of interplay with the Fabry Perot fringes. The left scale is the waveguide transmission and the right scale is the signal detected by the IR camera.}
\label{fig:exp_setup_sample}
\end{figure}

The photonic crystal structure studied here is the well-known optimized three missing holes (L3) PC microcavity \cite{akahane_03} in an air slab structure (thickness is 265 nm) based on a triangular lattice (period $a=400nm$) of holes with radius $r=0.24a$. The holes at the cavity edge were shifted by $s~=~0.15~a$. The cavity is side-coupled (the spacing is 3 rows) to a 1 mm long line-defect waveguide along the $\Gamma K$ direction, with width $W=1.05\sqrt{3}~a$ (see figure \ref{fig:exp_setup_sample}). The fabrication process and the detailed linear characterization of similar structures are described elsewhere \cite{combrie_05}. The loaded and intrinsic Q-factors,  7,000 and 30,000, respectively, are estimated from measurements using the procedure discussed in a previous paper \cite{combrie_opex06}.  The cavity resonant wavelength is 1567 nm. The characterisation setup consist of a tunable external cavity semiconductor laser (Tunics) with a relative accuracy of $1 pm$ and narrow linewidth ($<<$ 1 MHz), which is amplified by an Erbium Doped Fiber Amplifier (EDFA). The polarisation is set to TE, or electric field in the plane of the slab. Coupling to the PC waveguide is obtained through microscope objectives (Zeiss, N.A. 0.95) and micropositioners. The transmitted signal is detected with an InGaAs photodiode (bandwidth = 2 GHz), amplified (40dB, transimpedance, 2GHz bandwidth) and monitored by an oscilloscope (Tektronix 12.5 GS/s).

Two distinct regimes of modulation are considered: 1) sinusoidal modulation (frequency from few kHz to 50 MHz) and 2) low duty cycle (20 ns) square pulses.  The cavity is observed from the top with a microscope (long working distance objective Zeiss, N.A. 0.4) and an Infrared  InGaAs camera (Xenics). When the cavity is on resonance a bright spot is observed.

\subsection{Sinusoidal modulation}
 
\begin{figure}[h]
\includegraphics[width=8.5 cm]{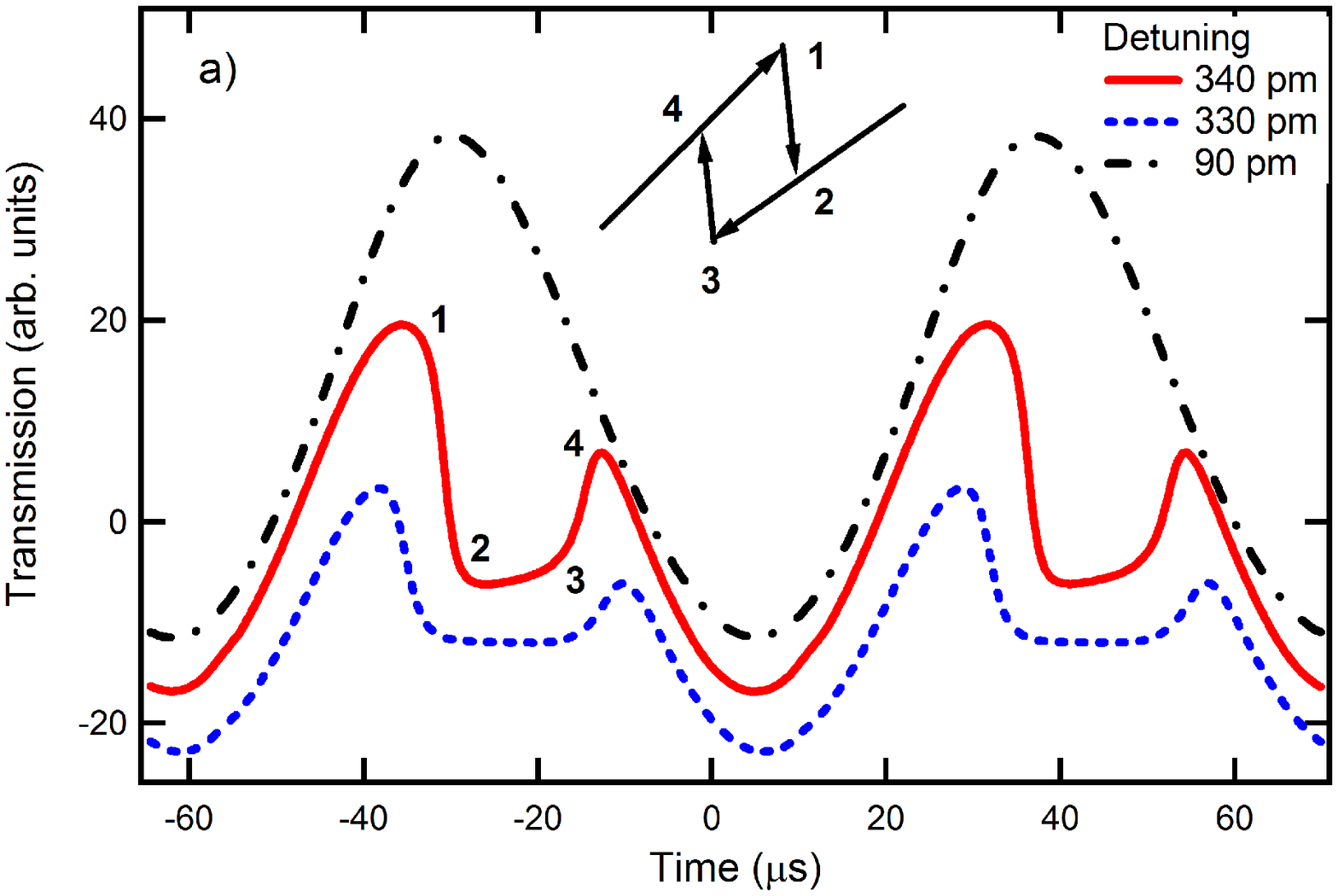} 
\includegraphics[width=7.5 cm]{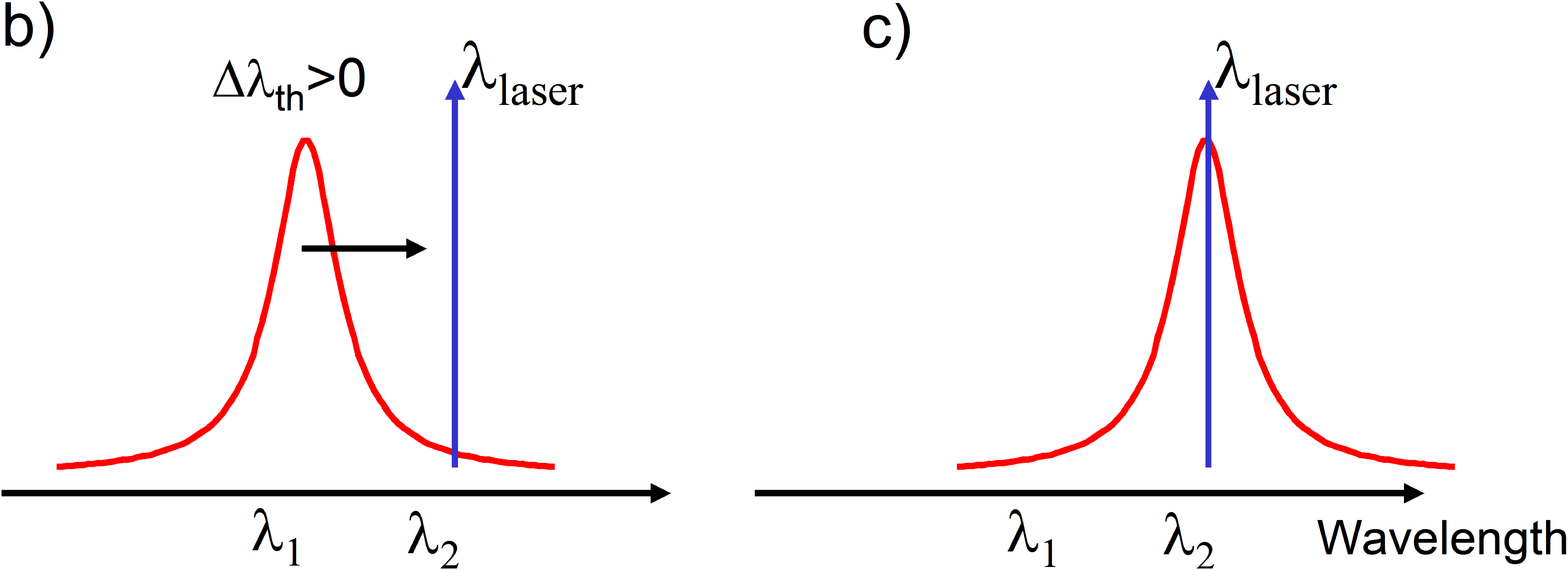} 
\caption{(Color online) Low modulation frequency experiment. Oscilloscope traces (a) showing the typical bistable response of a PC microcavity at low modulation frequency (15 kHz). The average input power in the PC waveguide is about 80 $\mu$W. The bistable cycle is also sketched and the transitions 1-4 are also marked. Detunings are +0.4, +1.5 and +1.55$\Delta\lambda_{FWHM}$ respectively. Scheme representing a cavity with a negative detuning b) and going to resonance due to nonlinearly induced red-shift c).}
\label{fig:bistable_low_freq}  
\end{figure}

At a low modulation rate (kHz) it has been shown by several groups\cite{notomi05,tanabe_ox05,Yang_07,weidner07,combrie2008} that bistability occurs at a fairly low optical power ($\mu W$ range, coupled into the waveguide). The dominant nonlinear effect is thermal induced index change because of heating resulting from carriers generated by TPA. Since this effect leads to a red-shift of the cavity frequency, bistability is observed when the initial detuning  $\Delta\lambda|_{t=0} =\lambda_{laser}-\lambda_{cavity}|_{t=0}$) is positive, where $\lambda_{cavity}|_{t=0}$ is the cold cavity resonant wavelength. This is shown in fig. \ref{fig:bistable_low_freq}. Typically, the detuning is set between $\sqrt{3}/2$ (the theoretical minimum, although it has been shown that the Fabry-P\'erot resonances in the waveguide affect this value \cite{Yang_07}) and several times  the cavity linewidth, estimated to be $\Delta\lambda_{FWHM} = 220 $ pm in our case. If the modulation frequency is increased while the detuning and the average signal power are kept constant, the bistable behavior disappears. For instance, with a detuning $\delta\lambda \approx 330 pm \approx 1.5 \Delta\lambda_{FWHM}$ and a modulation frequency of 15 kHz, bistability is observed in this sample at $80 \mu W$.  The power coupled into the waveguide is estimated considering the amplified laser power and the input objective coupling factor (here $\sim$ -10 dB) as in our previous work \cite{weidner07}.

\begin{figure}[h]
\centering
\includegraphics[width=8.5cm]{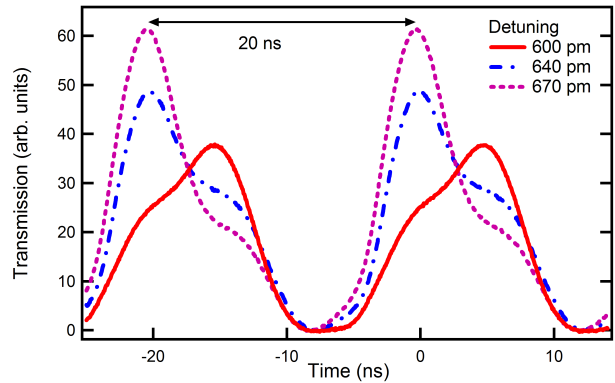}
\caption{(Color online) High modulation frequency experiment: oscilloscope traces of the transmitted signal with sinusoidal modulation (50 MHz)  for different detunings. The average power is about 2.5 mW.}
\label{fig:sin_modulation} 
\end{figure}

When the modulation frequency is increased to 1 MHz, the bistable effect and all traces of nonlinear distortion of the transmitted signal disappear. This could be explained in terms of the slow response of the thermo-optical effect. An important question arises whether the carrier plasma effect, much faster but weaker, will then take over. To answer that, the signal power and the modulation frequency were increased further and the detuning (still positive) was swept continuously from zero to about +700 pm, with the laser always on. This detuning value is larger than in low frequency experiments. A strong nonlinear distortion of the transmitted signal is then observed.
The onset of the nonlinear behavior is simultaneous to the observation of a bright spot through the IR camera, thus indicating that the resonant frequency of the cavity is red-shifted under a sinusoidal modulation pump. In particular, fig. \ref{fig:sin_modulation} shows a typical output signal when the modulation frequency is increased to $50$ MHz and the average power coupled in the waveguide is about 2.5 mW. The thermo-optic effect, driving the bistability at low modulation frequency, integrates over time. This explains ultra-low power levels. At higher modulation frequencies the power required to observe nonlinear effects increases. The nonlinear behavior is observed with a positive detuning from 600 pm and 670 pm ($\approx 3\Delta\lambda_{FWHM}$).
\indent We don't believe that the observed nonlinear distortion corresponds to a bistable behavior. Indeed, a fast nonlinear mechanism (plasma induced index change) should be dominate all other nonlinearities. That is probably not the case, as the short lifetime of carriers in GaAs compared to silicon drastically reduces the strength of this effect, compared to thermo-optic effects.

The nonlinear distortion observed in figure \ref{fig:sin_modulation} is quite general and can be reproduced at different modulation frequencies in the 10 - 100 MHz range. The nonlinear transitions are not sharp, compared to the modulation period (response time is a few ns). This time is however much faster than what could be attributed to a thermal effect. Additionally, the period, T = 20 ns, is much shorter than what is estimated to be the thermal recovery time in air slab PC cavities  \cite{notomi05}. We claim that the faster carrier plasma effect plays a key role at the ns time scale, while the thermal effect accumulates to strongly redshift the cavity.  
\begin{figure}[h]
\centering
\includegraphics[width=7 cm]{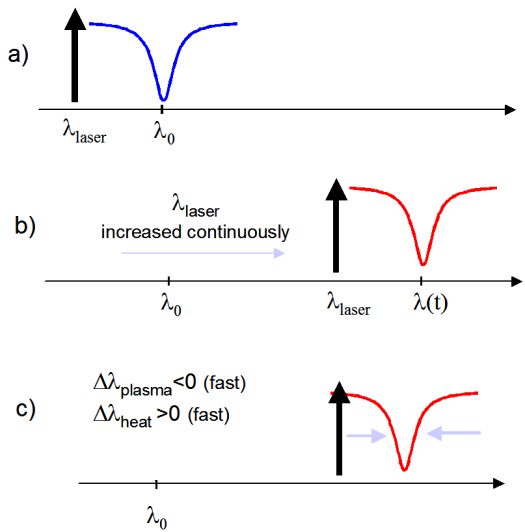} 
\caption{(Color online)  Scheme representing the interplay of thermal and carrier plasma effect and how they are related to the detuning. The system is at rest (cavity at its cold frequency) and the laser is switched on (a). The laser wavelength is increased but heating keeps the cavity red detuned (b). When the cavity temperature is high enough, the mean resonant wavelength get closer to the laser and the nonlinear distortion is apparent (c).}
\label{fig:scheme_detuning} 
\end{figure}

This mechanism is represented in Fig. \ref{fig:scheme_detuning}. Initially (a), the cavity is at room temperature and laser is detuned negatively ($\Delta\lambda<0$). The laser wavelength is increased continuously (b). The cavity is blueshifted (plasma effect) very fast, but heating quickly dominates and keeps the cavity redshifted until the end of the modulation cycle. Therefore, the cavity will be  pushed towards longer wavelengths as long as the heat generated by TPA over each cycle compensates the thermal flow across the membrane. Over each modulation cycle, the cavity resonant wavelength will oscillate about an equilibrium point set by this thermal balance alternating blue-shift and fast heating (c). The strongest distortion of the sinusoidal modulation is observed nearest to the largest detuning which can be sustained. In our case, we achieved detunings of up to 700 pm, which is related to a local temperature increase of 6 K through the thermo-optical coefficient (table. \ref{tab:physical_values}). Better physical insight into the two competing effects (thermal and free carrier), is gained by reducing the average input power, e.g considering much smaller duty cycles.   

\subsection{Low duty cycle excitation}

\begin{figure}[h]
\centering
\includegraphics[width=9cm]{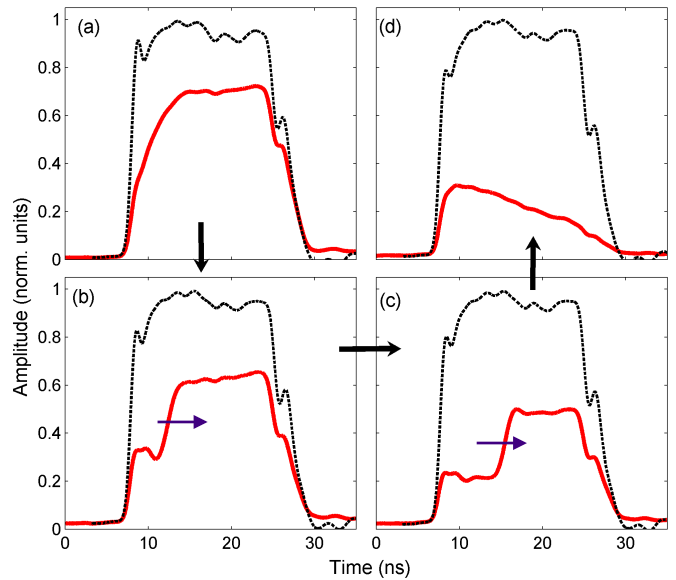}  
\caption{ (Color online) Experiment with \textquotedblleft non heating" pulses. Input (dashed) and output (solid) pulses as a function of the detuning. (a) $\Delta\lambda=-0.5\Delta\lambda_{FWHM}$, (b) $+0.7\Delta\lambda_{FWHM}$, (c) $+0.9\Delta\lambda_{FWHM}$ and (d) $+1.5\Delta\lambda_{FWHM}$. The pulse period is $2\mu s$.}
\label{fig:exp_pulse} 
\end{figure}

The cavity is excited with low duty cycle, square pulses (duration  $\Delta$t = 20 ns, period T=2 $\mu$s), so that the cavity has time to cool down before the arrival of the next pulse. We refer to this case as a \textquotedblleft non-heating pulse" because the average temperature of the cavity remains close to room temperature. Fig. \ref{fig:exp_pulse} reports the input and the transmitted pulse depending at various wavelength detunings. 
The input power in the pulse is estimated to be on the order of 10 mW. A more accurate determination of this value was complicated by the very low filling factor and the use of the EDFA, which is not designed to operate in these conditions. This lead to some fluctuation in the level of the amplified pulse. Furthermore, very low duty cycle means that the the power contained in the pulse is a fraction of the amplified spontaneous emission(ASE), which means that it is difficult to determine the exact value of the measurement of the average power.
\indent Let us first consider a negative detuning: e.g. $\Delta\lambda \approx -0.5\Delta\lambda_{FWHM}$ (fig. \ref{fig:exp_pulse}a). The leading edge of the output pulse rises with almost linear slope and is about 5 ns long. For zero or moderate positive detuning (e.g. $\Delta\lambda$ between 0 and $\approx \Delta\lambda_{FWHM}$), the output pulse typically has a two step shape with the leading part of the pulse having a lower level (fig. \ref{fig:exp_pulse}b and c). The transition between the two levels is sharp (a few ns) and moves from the leading edge to the trailing edge of the pulse as the detuning is increased. If detuning is increased further, the pulse takes on an almost triangular shape with a sharp leading edge and a long trailing edge (figure \ref{fig:exp_pulse} d) before losing any signature of nonlinear distortion.

Interestingly, the initial leading edge of the output pulse is very similar in the four cases considered in fig. \ref{fig:exp_pulse}, i.e. we don't see low transmittance associated to the on-resonance case, as long as the power is high. We see this behavior at much lower power only. We believe that this is due to a fast carrier effect which detunes the cavity faster than our detection apparatus is able to measure (1 ns).

It is also important to note that no signature of a plasma-induced bistable state is observed when the detuning is negative. We think that this is because of fast heating, which dominates the plasma effect before the pulse is extinguished. These features are well explained by our model.

\section{Nonlinear dynamical model}
The coupled-mode model developed here is based on previous optical microcavity literature. Particular structures that have been investigated include: microdisks \cite{Carmon2004} and photonic crystal cavities \cite{Barclay05,Uesugi06,cmtRaman}. The dynamical variables are: the optical energy inside the cavity $|a(t)|^2$, the free carrier density $N(t)$ and the cavity and membrane temperature $T(t)$. The model considers a single mode \footnote{the \textit{L3} cavity considered here is single mode over a large spectral range} cavity (resonance is at $\omega_0$ and unloaded quality factor $Q_0$), which is side-coupled to a waveguide (the loaded Q-factor is $Q$).
The modulation of the transmission observed in fig.\ref{fig:exp_setup_sample} is related to the Fabry-Perot resonance due to the finite reflectivity at the waveguide end facets. This spatially extended resonance has no impact on the nonlinear response (other than merely modulating spectrally the coupling into the cavity), since the associated field intensity is orders of magniture smaller than in the cavity.
The field in the cavity follows the equation:
\begin{equation}
\label{cmt_field}
\frac{\partial a(t)}{\partial t} =   ( i \omega - i\omega_L - \frac{\Gamma_{tot}}{2}) a(t) +\sqrt{\frac{\Gamma_c}{2}P_{in}(t) }
\end{equation} 
where $\omega_L$ is the laser frequency, $\Gamma_c=\omega_0 (Q^{-1}-Q_0^{-1})$ gives the cavity to waveguide coupling strength, $\Gamma_{tot}$ is the inverse (instantaneous) cavity lifetime, $P_{in}(t)$ is the power in the waveguide and $\omega= \omega_0 + \Delta\omega_{NL} $ is the instantaneous cavity frequency. Following \cite{Barclay05,Uesugi06}, the nonlinear change of the cavity frequency is given by:
\begin{gather}\label{delta_omega_nl}
\begin{split}
\Delta\omega_{NL}&=-\frac{\omega}{n_{eff}} \Delta n = -\frac{\omega}{n_{eff}} [\frac{n_{2I}\,c}{n\,V_{Kerr}} |a(t)|^2
\\ & \frac{dn}{dT}\Delta T(t) + \frac{dn}{dN}N(t)] 
\end{split}
\end{gather} 
Here $n$ is the refractive index of the bulk material and $n_{eff}$ the effective refractive index in the cavity, i.e.: $n_{eff}^2=\int n(\vec{r})^2\left|E(\vec{r})\right|^2dV/\int\left|E(\vec{r})\right|^2dV$,  and  $n_{2I}$ is the Kerr coefficient and $V_{Kerr}$ the Kerr nonlinear volume, defined as:

\begin{equation}
\label{eq_Kvolume} 
V_{Kerr}^{-1}=\frac{\int{n_{2I}(\vec r)/n_{2I}\left|E(\vec r)\right|^4 n(\vec r)^4 dV}}{(\int{n(\vec r )^2 \left|E(\vec r)\right|^2 dV})^2}
\end{equation}

The refractive index change due to the plasma effect is $dn/dN = -\omega_p^2/2n\omega^2N$, with $\omega_p^2=e^2N/\epsilon_0m^*$ the plasma frequency. In principle, holes and electrons both contribute to this effect, however, given the much smaller effective mass of electrons in GaAs, the contribution of holes is negligible. The inverse instantaneous photon lifetime is:
\begin{equation}
\label{gamma_tot}
\Gamma_{tot}= \frac{\omega_0}{Q} +  \Gamma_{TPA} + \Gamma_{FCA}
\end{equation} 
the first term is the inverse linear cavity lifetime, $\Gamma_{TPA}$ and $\Gamma_{FCA}$ are the contributions from two photon (TPA) and free carrier absorption (FCA), respectively. The TPA term is: $ \Gamma_{TPA} = \beta_2 c^2/n^2 |a(t)|^2/V_{TPA} $ with $\beta_2$ representing the TPA coefficient in units of $m/W$ and $V_{TPA}=V_{Kerr}$ the nonlinear effective volume\cite{Uesugi06}, while the free carrier absorption is proportional to the combined free carrier density $\Gamma_{FCA}=(\sigma_{e}+\sigma_{h}) N(t) c/n $. The evolution of carrier density follows the rate equation:
\begin{equation}
\label{carrier_rate}
\frac{\partial N(t)}{\partial t} = -\frac{N(t)}{\tau_N} + \frac{c^2/n^2 \beta_2}{V_{TPA} } \frac{1}{2\hbar \omega} \frac{1}{V_{car}} |a|^4 
\end{equation}
Here  $\hbar$ is the reduced Planck's constant, $\tau_{N}$ is the effective carrier lifetime, and $V_{car}$ is the volume in which the carriers spread and recombine\cite{Uesugi06}). In this approximation, the population decay is dominated by recombination at the surface, due to the large surface to volume ratio typical of photonic crystal strucures. Carriers are assumed to spread and distribute homogeneously within the carrier volume $V_{car}$, which is assumed to correspond to the region of the membrane delimited by the holes around the cavity. This approximation is rough, but it is a reasonable choice.
\begin{figure}[h]
\centering
\includegraphics[width=9cm]{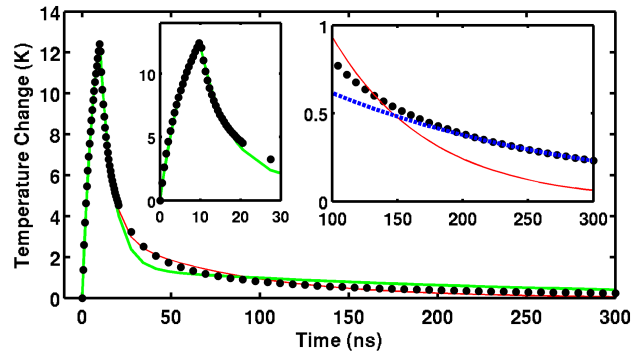} 
\caption{(Color online)  Modelling heat diffusion in the PC membrane. Local cavity temperature $T_c$ vs. time calculated with a 2D thermal diffusion equation (dots) and fitting (solid thick and thin lines) with the model eq. \ref{thermal_rate}. Inset: enlarged view of the short and the long term behaviour.  Inset right: fitted $\tau_{th,m} = 200ns$ with a single exponential (dashed line).Fit with 4 unconstrained parameters (thin line). Constrained fit with $\tau_{th,m} = 200ns$ $\tau_{th,c}=8.4\,ns$ (thick solid line, main plot and inset left).
Thermal capacitances for the cavity and the membrane are $0.43\,10^{-12} W/K$ and $3.4\,10^{-12} W/K$ respectively.} 
\label{fig:ThermalDiffusion} 
\end{figure}

\indent Because of the high thermal resistance in membrane PCs, thermal effects are very important and must be modelled appropriately. This is, for instance, the scope of ref.\cite{Kawashima_JQE2008}.  In contrast with existing literature, we introduce a more complicated model for the thermal response. We assume that the heat generated in the cavity has not yet diffused to the border of the membrane at the time scale of interest (1 ns). Therefore, a very important physical quantity is the thermal capacitance of the cavity $C_{th,c}=c_v\rho V_{th,c}$, associated with a small region of the membrane, with volume  $V_{th,c}$ roughly corresponding to the cavity volume.  
Heating and the spreading of the heat are modelled by solving the two dimensional heat diffusion equation. The heat is generated for some time (10 ns) inside the cavity and then the system is allowed to cool down. We assume that the radiative and convective contributions are negligible; therefore, since the membrane is suspended in air, all the heat has to flow through it. The result of this calculation is shown in fig. \ref{fig:ThermalDiffusion}. The cavity temperature $T_c$ increases with a rate that is governed by $C_{th,c}$  and $T_c$ decreases at two exponential time scales. The fast time scale $\tau_{th,c}$ is associated to a relatively fast transfer of heat from the cavity, where it is generated, to the neighbouring region in the membrane. We associate a second thermal capacitance $C_{th,m}$ to the whole membrane. The second time scale $\tau_{th,m}$ takes into account the spreading of the heat over the rest of the membrane (with temperature $T_m$) to the bulk semiconductor structure.
This picture is substantially different from previous models with a single time constant  $\tau=C_{th,c}R_{th}$ that is obtained from the thermal resistance $R_{th}$, defined as $W=R_{th}\Delta T$ (ratio between the increase of the temperature over the heating power at steady state). We will show that these two time scales play a crucial role in understanding the system response under sinusoidal and single pulse excitation.
The corresponding model entails therefore two auxiliary equations:


\begin{eqnarray}
\label{thermal_rate}
\frac{\partial T_c(t)}{\partial t} = -\frac{T_c(t)-T_m(t)}{\tau_{th,c}} +\frac{2 \hbar \omega}{\tau_N} \frac{V_{car} }{C_{th,c} V_{th}}N(t) \\
\frac{\partial T_m(t)}{\partial t} = -\frac{T_m(t)-T_{0}}{\tau_{th,m}}  -\frac{T_1(t)-T_2}{\tau_{th,1}}\frac{C_{th,c}}{C_{th,m}}
\end{eqnarray}

The thermal effective volume of the cavity and of the membrane $V_{th,c}$ $V_{th,m}$ and their thermal relaxation times $\tau_{th,c}$, $\tau_{th,m}$ are obtained by fitting the solution of the heat diffusion equation with the solution of the two time constant model (eq. \ref{thermal_rate}). If we fit all four parameters (two time constants and two capacitances) the fit turns out to be very good in the first 50 ns, yet tends to underestimate the long time constant and therefore to underestimate the total thermal resistance. To prevent this, we first fit a simple exponential to the long term behaviour, which gives a time constant of approximately 200ns. Then we fit the complete curve by constraining $\tau_{th,m} = 200 ns$. The result is shown in fig. \ref{fig:ThermalDiffusion} and confirms that this model is well adapted to describe the thermal behaviour of PC microcavities. The result of this calculation is given in table \ref{tab:physical_values}.


\begin{table}
\caption{\label{tab:physical_values} Physical paramenters used in the dynamical model.}
\begin{ruledtabular}
\begin{tabular}{llll}
Parameter&Symbol&Value&Ref.\\ 
\hline
\\
TPA coefficient    & $\beta_2 (cm/GW)$   &  $10.2$  & \cite{Dinu2003}\\
Kerr coefficient   & $n_{2I} (cm^2/W)$ &  $1.6 10^{-13}$  & \cite{Dinu2003} \\
Loaded Q   & $Q$ &  $7000$  &  meas.\\
Intrinsic Q   & $Q_0$ &  $30000$  &  est.\\
Modal Volume   & $V_{mod}$ &  $0.66 (\lambda/n_0)^3$  &  calc.\\
TPA Volume   & $V_{TPA} $ &  $3.13 (\lambda/n_0)^3$  &  calc.\\
Carrier Volume   & $V_{car} $ &  $7~V_{mod}$  &  \\ 
Thermo-optic coeff. & $dn/dT$ ($K^{-1})$ & $2.48~10^{-4}$  & \cite{handook_optics} \\
Therm. eff. vol.(c)   & $V_{th,c} $ &  $3.8*V_{mod}$& calc. \\ 
Therm. eff. vol. (m)   & $V_{th,m} $ &  $31*V_{mod}$& calc. \\
Specific Heat		&	$c_v\rho$ ($W/K\,m^{-3}$)    &  1.84 $10^6$  & \cite{handook_optics} \\
Carrier lifetime   & $\tau_{N} $ &   8 ps  & \cite{bristow_GaAs_APL03,Husko_APL2008}\\ 
Therm. relax. time (c)   & $\tau_{th,c} $ &  8.5 ns & calc. \\
Therm. relax. time (m)  & $\tau_{th,m} $ &  200 ns  & calc.\\
Therm. resistance  & $R_{th}=\sum\frac{\tau_{th}}{C_{th}} $ &  $7.5\,10^4$ K/W & calc.\\
FCA cross section & $ \sigma_{e,h}$  $(10^{-22} m^2)$ 	 &  $9.3 (\frac{\lambda}{1.0\mu m})^{2.3}$ & \cite{Reinhart_JAP05}\\
\end{tabular}
\end{ruledtabular}
\end{table}

\section{Discussion}
Simulations have been carried out with parameters given in Table \ref{tab:physical_values}. All of them are well known or can be calculated or measured with reasonable accuracy. The carrier volume $V_{car}$ is the exception. The impact of carrier diffusion has been investigated theoretically in a recent paper \cite{Tanabe_JLT2008}, providing some hints to explain a fast recovery time in Silicon PCs. The carrier lifetime in patterned GaAs (e.g. 2D PCs) is however much shorter than in Silicon. Very recently we estimated it to be about 6 ps in our GaAs cavities\cite{Husko_APL2008}, which is consistent with previous (8 ps) estimates for GaAs PC structures \cite{bristow_GaAs_APL03}. 

\indent In the limit where the dynamic is much slower than $\tau_N$, which the case considered in this work, $\tau_N$ and $V_{car}$ play the same role and what matters here is the ratio $\tau_N/V_{car}$. To show that, let us consider the density of the generated carriers : 
\begin{equation}
\label{carriers_steady}
N(t)= \frac{c^2/n^2 \beta_2}{V_{TPA} } \frac{1}{2\hbar \omega} \frac{\tau_N}{V_{car}} |a|^4 
\end{equation}
The blue-shift of the cavity resonance due to the generated carriers is therefore proportional to the instantaneous power absorbed, $P_a$. The red shift induced by heating is proportional to the absorbed energy and to the inverse of the thermal capacitance. Then, the relative strength of thermal and carrier effects depends on the ratio $\tau_N/V_{car}$ and on the thermal capacitance. The dynamical behaviour of the system investigated here is basically determined by these two quantities. 

Modelling was performed by varying only one parameter, the carrier volume, and adjusting the input power and the detuning around reasonable values. All other values are well known or calculated precisely. 
\begin{figure}[h]
\centering
\includegraphics[width=9cm]{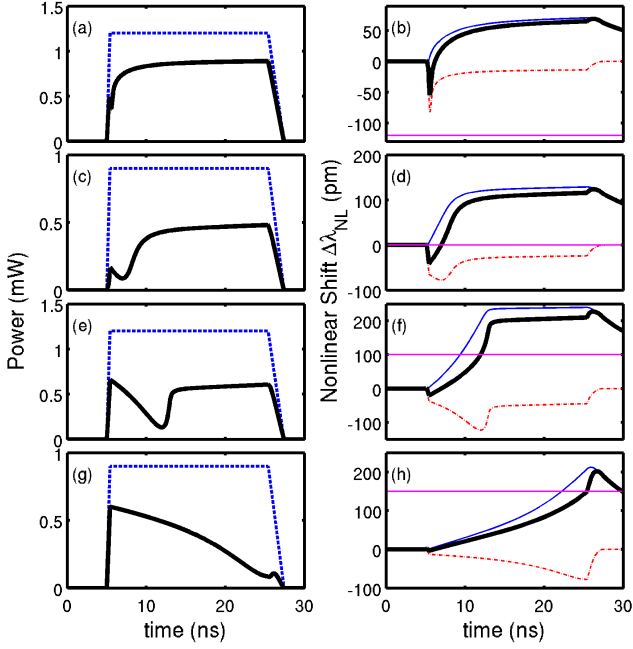} 
\caption{(Color online) Simulation of the response (solid) to a single pulse (dashed), depending on the detuning $\Delta\lambda(0)$: -120 pm (a,b), 0 pm (c,d), 80 pm (e,f) and 150 pm (e,f). Right: corresponding instantaneous frequency shift $\Delta\lambda_{NL}$ of the cavity resonance (solid line) and carrier plasma (thin-dashed) and thermo-optical contributions (dotted). The laser frequency is also shown (thin solid). $P_{in}$ = 1.2 mW, 0.9 mW, 1.2 mW and  1.0 mW respectively.}
\label{fig:fig_simulazione_pulse} 
\end{figure}

\begin{figure}[h]
\centering
\includegraphics[width=9cm]{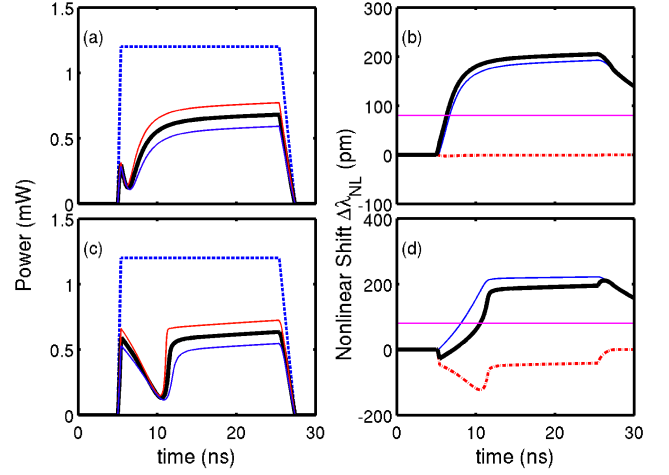} 
\caption{(Color online)  Simulated response when the contribution to the plasma shift from carrier plasma is suppressed (a) and full model (c). The excitation power is 1.2 mW (thick line). The case with $P_{in}$ =1.1 mW and 1.3 mW are also plot (thin lines). The detuning $\Delta\lambda(0)$ is 80 pm. The corresponding instantaneous frequency shifts  $\Delta\lambda_{NL}$ is also plot (b,d) for the case $P_{in}$ =1.2 mW.}
\label{fig:fig_simulazione_withwo_plasma} 
\end{figure}
\indent Fig. \ref{fig:fig_simulazione_pulse} reports the simulated responses of a square pulse at various detunings. When the initial detuning is negative (e.g. $\Delta\lambda(0)=-120 pm$, fig.  \ref{fig:fig_simulazione_pulse} a and b), the carrier effect is responsible for the drop of the leading edge of the transmitted pulse. The signal coupled in the cavity is strong enough to cause considerable generation of carriers and the resulting blue-shift tends to tune the cavity into resonance so that transmission is reduced (the cavity is side-coupled).
Within a ns time scale, the heating following carrier recombination red-shifts and therefore detunes the cavity again. Such fast heating arises from the small thermal capacitance of the cavity. After the pulse is extinguished, the cavity is red-shifted by about 100 pm with respect to the initial position and recovers its initial state in a fraction of $\mu s$. 
\indent Let us now consider the case of zero detuning (fig.  \ref{fig:fig_simulazione_pulse}c and d). The carrier induced index change produces an instantaneous (with respect to the time scale considered here) blue shift that adds to the initial detuning. Heating, following absorption, tunes the cavity back and transmission decreases once again. Thus more carriers are generated and the plasma effect tends to oppose heating but, in the end, the heating dominates. 
At these modulation rates thermal relaxation is not sufficiently fast to dissipate heat from the cavity after each duty cycle. Thus heat accumulates over time, causing a net red-shift of the cavity resonance. When the instantaneous detuning $\Delta\lambda_{NL}$ becomes positive, carrier generation starts to decrease and so does the carrier induced shift. This initiates a positive feedback that quickly detunes (red) the cavity and makes the carrier density drop very fast. The delay between the pulse leading edge and this transition is clearly related to the amount of the initial detuning (fig. \ref{fig:fig_simulazione_pulse}e and f). 
When the initial detuning is positive and large enough (fig. \ref{fig:fig_simulazione_pulse}g and h), then heating tends to tune the cavity into resonance, thus reducing the transmission but also generating carriers. The plasma effect will partially oppose the red-shift, thus explaining an almost linear change with time.
\indent The sharp step observed in fig. \ref{fig:exp_pulse}b,c can only be explained by the thermal effect  overtaking the carrier plasma index change. This step cannot be reproduced for any choice of free parameters if the plasma induced index change is suppressed in the model. This is shown in figure \ref{fig:fig_simulazione_withwo_plasma}. In panel a and b we ran the same simulation as in fig.  \ref{fig:fig_simulazione_pulse}e, but we have suppressed the contribution of carriers to the frequency shift of the cavity. In this case the transmission increases with a linear, but finite, slope after the dip, independent of the pulse power. This is not what we see experimentally. Conversely, when the plasma effect is included, the sharp transition is reproduced correctly.
There is a physical reason for that: the observed steep step requires a fast index change with, say, a negative sign, and a slower index change, with opposite sign. A simple explanation is that at first the plasma effect dominates (over a very short time, a few ns) before thermal heating takes over. If a plasma effect is ruled out, there is no way to explain the experimental results. A Kerr effect cannot be considered because it has the same sign as the thermal effect.
After the submission of this manuscript, critical slowing down (CSD), implying a transition from a low (off-resonance) to a high (on-resonance) transmission state, has been reported in InP-based PC cavities\cite{Yosia2008}. This effect also results into a step in the pulse response. We believe however that this is a different mechanism than what we describe here. The reason is that our system is side-coupled, thus CSD would manifest as a transition from a high  (off-resonance) to a low (on resonance) transmission state. We did not observe that. A possible explanation is the carrier lifetime  in GaAs PCs being mush shorter than in Silicon or InP PCs. This implies that the plasma dispersion effect is much weaker than the thermal induced index change.

\indent We found experimentally that the response of the cavity is very sensitive to the power of the pulse, in particular in the case reported in \ref{fig:exp_pulse}e. This behavior is well reproduced by theory. This means that input power must be set precisely in simulation and it cannot be the same for all the detunings considered here.
Indeed, experimentally the peak pulse power is not constant since, a) the output power from the amplifier showed a marked dependence on slight change in wavelength, b) the waveguide transmission is modulated by a strong Fabry-P\'erot effect (fig. \ref{fig:exp_setup_sample}).
The fact that the detuning in experiments and in the theory are not the same can be understood by accepting some error in the measurement of the detuning.  The reason is that the laser was directly modulated which induced some deviation from the nominal setting of the laser on the order of 100 pm.

\begin{figure}
\centering
\includegraphics[width=9cm]{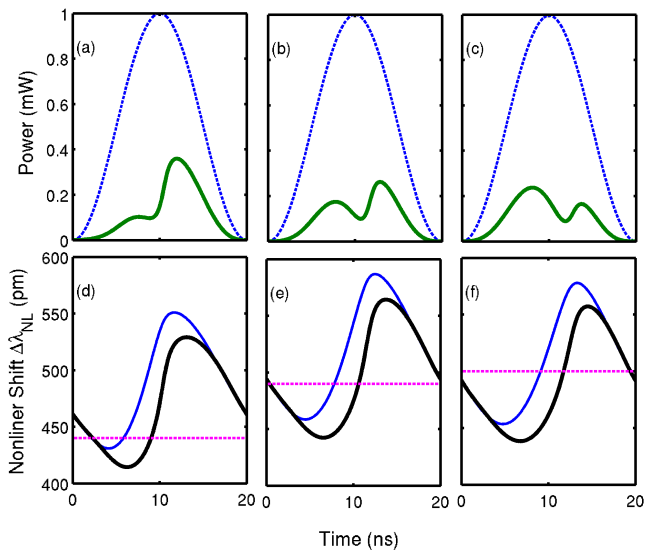}
\caption{(Color online) Simulation of the response of the cavity to a sinusoidal excitation for the following detunings (from top to bottom): 440 pm (a,d), 490 pm (b,e) and 500 pm (c,f). Input peak power is 1 mW. The transmitted pulses are shown on the top (a,b,c), the instantaneous frequency shift (solid think curve) and the thermo-optical contribution alone are shown on the bottom (d,e,f).}
\label{fig:sim_cw_ENT} 
\end{figure}

This relatively fast (few ns rise-time) nonlinear effect reported in fig. \ref{fig:exp_pulse} is well understood through modelling. The complexity of the model is justified by the number of different time scales present.
An important point of this paper is to demonstrate that the same model is able to explain also the response to a sinusoidal excitation. The response is calculated for a time long enough to ensure that we reach the steady-state regime. The simulations shown in fig. \ref{fig:sim_cw_ENT} are carried out in order to mimic experimental conditions, that is, the excitation wavelength is detuned negatively with respect to the cold cavity frequency. As the excitation is turned on, the wavelength is increased adiabatically with respect to the modulation period. Indeed it is found that the cavity stays red detuned, and is pushed towards longer wavelengths, until the excitation power and the heat generated is enough to sustain the detuning.

\indent The simulations clearly support the experimental results shown in figure \ref{fig:sin_modulation} and interpretation in fig. \ref{fig:scheme_detuning}. Indeed, if we now compare the measured and calculated oscilloscope traces we conclude that the dynamics are accurately reproduced, as the detunings used in modelling correspond to what is measured experimentally.

In particular, the response is very well understood by looking at the instantaneous frequency of the cavity, which crosses the laser frequency at different points depending on the detuning. If the detuning $\Delta\lambda$ is too small, the cavity wavelength is always higher than the laser frequency, as the instantaneous cavity nonlinear shift $\Delta\lambda_{NL}=\lambda_{cavity}(t)-\lambda_{cavity}(0)$ due to heating is much larger over the whole modulation cycle. When the detuning $\Delta\lambda$ increases, $\Delta\lambda_{NL}$ also increases on average, as the resonance is closer to the laser frequency, but less than $\Delta\lambda$ does. Thus, there exists a combination of pulse power, modulation frequency and detuning such that $\Delta\lambda$ and $\Delta\lambda_{NL}$ are close one to each other. If $\Delta\lambda$ is increased further, heating is not enough to keep $\Delta\lambda_{NL}$ close to it and nonlinear distortion is lost. The role of the plasma induced index change and the interplay with the thermo-optic effect is also clear in figure \ref{fig:fig_simulazione_pulse}(b,d,f,).  

It is interesting to note that the plasma effect vanishes when $t<4ns$ and $t>16ns$ in figs. \ref{fig:sim_cw_ENT}b,d and f,  although minima of the sinusoidal excitation are at t=0 and t=20 ns (figs. \ref{fig:sim_cw_ENT}a,c,e). This important point is understood when considering that while the plasma effect is almost instantaneous (within the time scale relevant to this work), the thermo-optic effect follows the time integral of the optical power. At time $t=0$ the carrier population is negligible and the cavity is cooling down, and therefore is approaching the laser wavelength. As the power inside the cavity increases, carriers are generated with a rate proportional to the power squared. A carrier induced index change builds up very fast and produces the difference between the thin and the thick curves. The power absorbed produces heating. Instantaneous heating is the time integral of instantaneous absorbed power and therefore builds up with a delay with respect to the plasma effect. The thermal effect then dominates the plasma effect. As soon as the resonance is crossed again and the detuning becomes comparable to half the cavity linewidth (100 pm), nonlinear absorption decreases very quickly and the carrier population declines accordingly. Therefore the carrier effect disappears. The cavity again starts to cool down and the cycle begins anew.

\section{Conclusion}
Switching based on TPA has a great potential in GaAs based structures, as it benefits from the strong nonlinear absorption coefficient $\beta_2$ and sub-ns carrier lifetimes typical of this material. However, most practical applications will require a high repetition rate (i.e. high duty cycle), therefore thermal effects are unavoidable in such small structures. In particular, heating is very strong in photonic crystals, because of the poor thermal conductivity of PC membranes. We have experimentally investigated this regime and confirmed results from other groups evidencing that the thermal dynamics in PC membrane cavities is much faster than in macroscopic photonic devices due to the small modal volume of the structure. More particularly, we have explored a regime in which a sharp (2 ns) transition result from a positive feedback between thermal and carrier induced refractive index changes. The peak power required to see these effects is at the mW level. We have also explained why it is not possible to observe bistability in GaAs cavities when the modulation rate is on the order of 10  - 100 MHz.
\indent Based on these experimental results, we have introduced a dynamical model for PC microcavities which includes two thermal relaxation constants in order to account for heat diffusion from the cavity to the neighbouring membrane. We show that this is crucial in order to understand the dynamical response of the cavity and that this model reproduces the experimental results quite well.
The ability to properly account for very different time scales is crucial for modelling patterning effects (i.e. long term or memory effects) which result not only from the carrier dynamics but also on fluctuations of the average power which induce changes in the local temperature of the cavity. This is crucial to design PC based all-optical switches for practical optical signals at high repetition rate (e.g. 10 GHz).

\indent One of the authors (C. Husko) thanks the Fulbright Grant for financial support.

\end{document}